\documentclass[traditabstract]{aa}

\pdfoutput=1
\usepackage{graphicx}
\usepackage[font=small]{caption}
\usepackage{indentfirst}
\usepackage{tabu}

\usepackage{natbib}
\usepackage{booktabs}
\usepackage{fixltx2e}

\usepackage{amsmath}
\usepackage{txfonts}
\usepackage{amssymb}
\usepackage[a]{esvect}
\usepackage{tikz}

\newcommand{\degree}[0]{$^{\circ}$}

\newcommand{\cbra}[1]{\left( #1 \right)}      

\let\baraccent=\=
\renewcommand{\=}[1]{\stackrel{#1}{=}} 
\let\arrowaccent=\>
\renewcommand{\>}[1]{\stackrel{#1}{\Rightarrow}} 

\makeatletter
\newcommand{\rmnum}[1]{{\footnotesize{\expandafter\@slowromancap\romannumeral #1@}}}
\newcommand{\Rmnum}[1]{{\expandafter\@slowromancap\romannumeral #1@}}
\makeatother

\everymath={\rm}
\newcommand{\tfm}[1]{\tablefootmark{#1}}
\begin{document}

\title{Star formation efficiency along the radio jet in Centaurus A}

\author{
   Q. Salom\'e\inst{1}  \and
   P. Salom\'e\inst{1}  \and
   F. Combes\inst{1,2}  \and
   S. Hamer\inst{1}     \and
   I. Heywood\inst{3,4}
}

\institute{
   LERMA, Observatoire de Paris, CNRS UMR 8112, 61 avenue de l'Observatoire, 75014 Paris, France \\ email: quentin.salome@obspm.fr \and
   Coll\`ege de France, 11 place Marcelin Berthelot, 75005 Paris \and
   CSIRO Astronomy \& Space Science, P.O. Box 76, Epping, NSW 1710, Australia \and
   Department of Physics \& Electronics, Rhodes University, P.O. Box 94, Grahamstown 6140, South Africa
}

\date{Received 25 April 2015 / Accepted 17 November 2015}

\titlerunning{SFE along the radio jet of Centaurus A}
\authorrunning{Salomé et al.}

\abstract{
   NGC 5128 (also known as Centaurus A) is the most nearby powerful AGN, widely studied at all wavelengths. Molecular gas has been found in the halo at a distance of $\sim 20\: kpc$ from the galaxy center, associated with H\rmnum{1} shells, through CO line detection at SEST \citep{Charmandaris_2000}. The molecular gas lies inside some IR and UV bright star-forming filaments that have recently been observed in the direction of the radio jets. These archival data from GALEX (FUV) and Herschel (IR) show that there is dust and very weak star formation (a few $10^{-5}-10^{-4}\: M_\odot.yr^{-1}$) on scales of hundreds of parsecs. NGC 5128 is thus a perfect target for detailed studies of the star formation processes at the interface of the jet/gas interaction. \\
   On top of analysing combined archival data, we have performed searches of HCN(1-0) and $HCO^+$(1-0) emission with ATCA at the interaction of the northern filaments and the northern H\rmnum{1} shell of Centaurus A. Measuring the dense gas is another indicator of star formation efficiency inside the filaments. However, we only derived upper limits $L'_{HCN}<1.6\times 10^3\: K.km.s^{-1}.pc^2$ and $L'_{HCO^+}<1.6\times 10^3\: K.km.s^{-1}.pc^2$ at $3\sigma$ in the synthesised beam of $3.1''$. Compared with the CO luminosity, this lead to a dense-to-molecular gas fraction $<23\%$. \\
   We also compared the CO masses with the SFR estimates in order to measure a star formation efficiency (SFE). Using a standard conversion factor leads to long depletion times (7 Gyr). We then corrected the mass estimates from metallicity effect by using gas-to-dust mass ratio as a proxy. From MUSE data, we estimated the metallicity spread ($0.4-0.8\: Z_\odot$) in an other region of the filament, that corresponds to gas-to-dust ratios of $\sim 200-400$. Assuming the same metallicity range in the CO-detected part of the filament, the CO/H$_2$ conversion ratio is corrected for low metallicity by a factor between 1.4 and 3.2. Such a low-metallicity correction leads to even more massive clouds with higher depletion times (16 Gyr). We finally present ALMA observations that detect 3 unresolved CO(2-1) clumps of size $<37\times 21\: pc$ and masses around $10^4\: M_\odot$. The velocity width of the CO emission line is $\sim 10\: km.s^{-1}$, leading to a rather high virial parameter. This is a hint of a turbulent gas probably powered by kinetic energy injection from the AGN jet/wind and leading to molecular gas reservoir not forming star efficiently. \\
   This work shows the importance of high resolution data analysis to bring a new light on the local processes of AGN/jet feedback likely negative (quenching star formation) in the case of Cen A filaments.}

\keywords{Methods:data analysis - Galaxies:evolution - interactions - star formation - Radio lines:galaxies}

\maketitle


\section{Introduction}

   Active galactic nuclei (AGN) are thought to play a role in galaxy evolution (and formation), but the exact effect they have is still debated. It is not yet clear how the so-called AGN feedback affects star formation. On the one hand, \textit{negative} feedback could prevent or regulate star formation through the energy released by the AGN (mechanical or radiative) (\citealt{Heckman_2014} and references therein). If this energy were transferred to the surrounding medium, it could either heat or expel the gas reservoir (e.g. \citealt{Fabian_2012,Heckman_2014} and references therein). \cite{Cicone_2014} observed a sample of ultraluminous infrared galaxies (ULIRG) with the Plateau de Bure interferometer and found massive gas outflows in four cases. The energy produced by the AGN was invoked to have blown out the cold gas from the galaxy and thus quench star formation. However, the details of this interaction between the AGN and the gas are still unclear. On the other hand, \cite{Zinn_2013} conducted a statistical study and showed that AGN with pronounced radio jets have a much higher star formation rate than those that do not. It is indeed expected that the propagation of jet-driven shocks can accelerate the gas cooling and trigger star formation \citep{Best_2012,Ivison_2012}, producing \textit{positive} AGN feedback. In local brightest cluster galaxies, evidence of radio-jet associations with star formation was observed by \cite{McNamara_1993}, while \cite{Emonts_2014} found CO-jet alignment in several high-z radio galaxies that they interpreted as the result of jet-induced star formation and gas cooling. Hydrodynamic simulations of radiative shock-cloud interactions show that a large portion of gas along the propagation direction of a shock may cool very efficiently \citep{Fragile_2004,Gaibler_2012}.

   Direct gas-jet interaction within the plane of galaxy discs is rare. The most famous example is NGC 4258 \citep{Herrnstein_1997} where the jet seems to have produced a cavity inside the disc gas and along the jet, and the optical line ratio shows evidence for bow shocks \citep{Cecil_2000}.

   \cite{Leroy_2008,Leroy_2013} compared the molecular gas content of nearby galaxies with star formation rate (SFR) tracers. With a resolution of $\sim 1\: kpc$, they studied the galaxies in detail. In spirals, they found that the total gas depletion time is roughly constant in the inner disc ($R_{gal}\lesssim R_{25}$) where gas is mostly molecular, and that it increases at larger radius where the atomic component dominates. For dwarf galaxies, they observed a similar behaviour. This result seems to indicate that the star formation efficiency depends mostly if not only on the amount molecular gas, as claimed by \cite{Schruba_2011}, who showed that the SFR remains correlated with $H_2$ in H\rmnum{1}-dominated environments.
\cite{Daddi_2010} and \cite{Genzel_2010} observed possible multiple "modes" of star formation depending on their efficiency for similar molecular gas surface densities. This would indicate that some environmental effects may in fact influence star formation.

   There is not much evidence of radio-jet and molecular gas interaction so far. 3C 285/09.6 \citep{vanBreugel_1993} and Minkowski's Object \citep{vanBreugel_1985}, two of the most famous examples were studied by \cite{SalomeQ_2015a}. These two star-forming regions lie along the jet direction, at distances of several kpc from the galaxy. Upper limits on the amount of molecular gas available to fuel star formation have been derived through CO non-detection. \cite{SalomeQ_2015a} used data obtained with the 30m telescope, which showed that both limits lie at least on or even above the Kennicutt-Schmidt (KS) law when using a standard CO/H$_2$ conversion factor. The molecular gas depletion time was found to be $<1\: Gyr$ in 3C285/09.6 and down to $<20\: Myr$ for Minkowski's Object in regions of $24''$ (i.e. $\sim 36\: kpc$ and $\sim 9\: kpc$). \\
We here focus on NGC 5128 (also known as Centaurus A). It is a giant nearby early-type galaxy (ETG) that lies at the heart of a moderately rich group of galaxies. It hosts a massive disc of dust, gas and young stars in its central regions \citep{Israel_1998}. This gas disc is misaligned in CO \citep{Espada_2009}. \cite{Davis_2011} observed a sample of 260 ETG from the $ATLAS^{3D}$ project and found that more than 40\% of the sample are kinematically misaligned, indicating that their gas was supplied from accretion and mergers. The star formation rate of the central galaxy has been estimated to $2\: M_\odot.yr^{-1}$ by \cite{Neff_2015b} from observations made with the Galaxy Evolution Explorer (GALEX) in the far-UV (FUV). The AGN at the centre of the galaxy is composed of radio jets that are about 1.35 kpc long and giant radio lobes that extend up to $\sim 250\: kpc$. NGC 5128 is surrounded by faint arc-like stellar shells (at a radius of several kpc around the galaxy). \cite{Schiminovich_1994} detected H\rmnum{1} gas in the shells, and \cite{Charmandaris_2000} observed CO emission at the intersection with the radio jet. In addition, \cite{Auld_2012} detected a large amount of dust ($\sim 10^5\: M_\odot$) around the northern shell region. They measured gas-to-dust and $H_\rmnum{1}/H_2$ ratios typical of late-type gas-rich galaxies. They concluded that both the dust and gas could come from a galaxy that merged with Cen A, forming the dusty disc at the centre.
\smallskip

   Along the radio-jets, optically bright blue filaments have been observed \citep{Mould_2000,Rejkuba_2001,Rejkuba_2002}, and from colour-magnitude diagrams identifying individual stellar populations, numerous OB-associations have been found, revealing jet-induced star formation, down to 10-15 Myr ago.  Some ionized gas does therefore originate from young stars, although there might exist a significant contribution from the AGN \citep{Morganti_1991,Santoro_2015b}. Some ionization could also come from shocks in the jet-ISM interaction (e.g. \citealt{Veilleux_1987,Dopita_1995,Dopita_1996}). UV emission is also present as detected by GALEX, and far infrared emission, although some part could be due to dust heating by old stars, especially at large scale \citep{Auld_2012}. \\
The so-called inner and outer filaments discovered by \cite{Morganti_1991} (see left panel of Fig. \ref{overview}) are located along the direction of the northern radio jet, at a distance of $\sim 7.7\: kpc$ and $\sim 13.5\: kpc$. The outer filament is larger ($\sim 8\: kpc$) than the inner filament ($\sim 2\: kpc$). \cite{Crockett_2012} interpreted the inner filament as the result of a weak cocoon-driven bow shock that propagates through the diffuse interstellar medium, triggering star formation. \cite{Kraft_2009} mapped huge X-ray filaments in the northern middle radio lobe. These very hot gas filaments could result from a jet-cloud interaction, where cold, dense clouds have been shock-heated to X-ray temperatures. Optical excitation lines have been observed with the Visible Multi Object Spectrograph (VIMOS) and the Multi Unit Spectroscopic Explorer (MUSE) in the inner \citep{Hamer_2015} and the outer filaments \citep{Santoro_2015a,Santoro_2015b}. Both filaments show distinct kinematical components: a well-defined knotty filament, and a more diffuse structure.
\cite{Neff_2015b} reported what they called a "weather ribbon" in the outer filament of Centaurus A. This ribbon is a northern extension of the optical emission-lines filament and is associated with a knotty ridge, young stars, emission-line clouds and diffuse radio emission. The authors hypothesised that the ribbon might be downstreaming ionised gas from the cold H\rmnum{1} shell, due to the interaction with the radio jet. This may support the scenario of jet-induced star formation because galactic winds can stimulate star formation in dense clouds. General properties of NGC 5128 and the outer filament are summarised in Table \ref{table:overview_CenA}.

\begin{table}[h]
  \centering
  \caption{\label{table:overview_CenA} General properties of Centaurus A and the northern outer filament \citep{Morganti_1991} at the intersection with the H\rmnum{1} shell \citep{Schiminovich_1994}.}
  \begin{tabu}{lcc}
    \hline \hline
    Source                                 &         Centaurus A         &       outer filaments       \\ \hline
    z,$v_{LSR}\: (km.s^{-1})$              &    0.001826, $\sim 545$     &          0.001826           \\
    distance (kpc)                         &              0              &     $\sim 13.5$ \tfm{a}     \\
    $D_A$,$D_L$ (Mpc)                      &             \multicolumn{2}{c}{3.42 \tfm{b}}              \\
    Scale (pc/$''$)                        &             \multicolumn{2}{c}{16.5 \tfm{b}}              \\ \hline
    RA (J2000)                             &     $13^h 25^m 27^s.6$      &     $13^h 26^m 18^s.9$      \\
    Dec (J2000)                            &        $-$43:01:09.8        &        $-$42:49:32.0        \\
    $L_{FUV}\: (erg.s^{-1})$               & $3.5\times 10^{42}$ \tfm{c} &     $3.2\times 10^{40}$     \\
    $SFR_{FUV}\: (M_\odot.yr^{-1})$        &        0.16 \tfm{d}         & $1.4\times 10^{-3}$ \tfm{d} \\
    $L_{FIR}\: (L_\odot)$                  &  $9.3\times 10^9$ \tfm{c}   &      $3.2\times 10^7$       \\
    $SFR_{FIR}\: (M_\odot.yr^{-1})$        &         1.6 \tfm{d}         & $5.5\times 10^{-3}$ \tfm{e} \\
    $L_{TIR}\: (L_\odot)$                  &              -              &      $3.3\times 10^7$       \\
    $SFR_{TIR}\: (M_\odot.yr^{-1})$        &              -              & $4.9\times 10^{-3}$ \tfm{d} \\ \hline
    $M_{H_2}\: (M_\odot)$                  &  $3.3\times 10^8$ \tfm{f}   &  $>1.7\times 10^7$ \tfm{g}  \\
    $t_{dep}\: (Gyr)$                      &             0.2             &           $>4.3$            \\ \hline
  \end{tabu}
  \tablefoot{
    \tablefoottext{a}{\cite{Morganti_1991}}
    \tablefoottext{b}{\cite{Ferrarese_2007}}
    \tablefoottext{c}{\cite{Neff_2015b}}
    \tablefoottext{d}{\cite{Kennicutt_2012}}
    \tablefoottext{e}{\cite{Kennicutt_1998}}
    \tablefoottext{f}{\cite{Eckart_1990}}
    \tablefoottext{g}{\cite{Charmandaris_2000}, in the shell S1} \\
    For the outer filament, all luminosities were estimated on a region of $10'\times 12'\sim 10\times12\: kpc$ (Fig. \ref{overview}) \citep{Auld_2012}. The FIR and TIR luminosity were computed on the 40-500 and $3-1100\: \mu m$ ranges, they do not contain a great contribution of old stars \citep{Rejkuba_2001} and are relevant tracers of star formation. The molecular gas mass and the molecular depletion time are derived for a fixed $\alpha_{CO}$.
  }
\end{table}

   In Sect. \ref{sec:Obs/res}, we present the data used for this study and the results we derived. Then we discuss these results in Sect. \ref{sec:discussion}. Throughout this paper, we assume the cold dark matter concordance Universe, with $H_0=70\: km.s^{-1}.Mpc^{-1}$, $\Omega_m=0.30$, and $\Omega_A=0.70$.

\begin{figure*}[h]
  \centering
  \includegraphics[width=\linewidth]{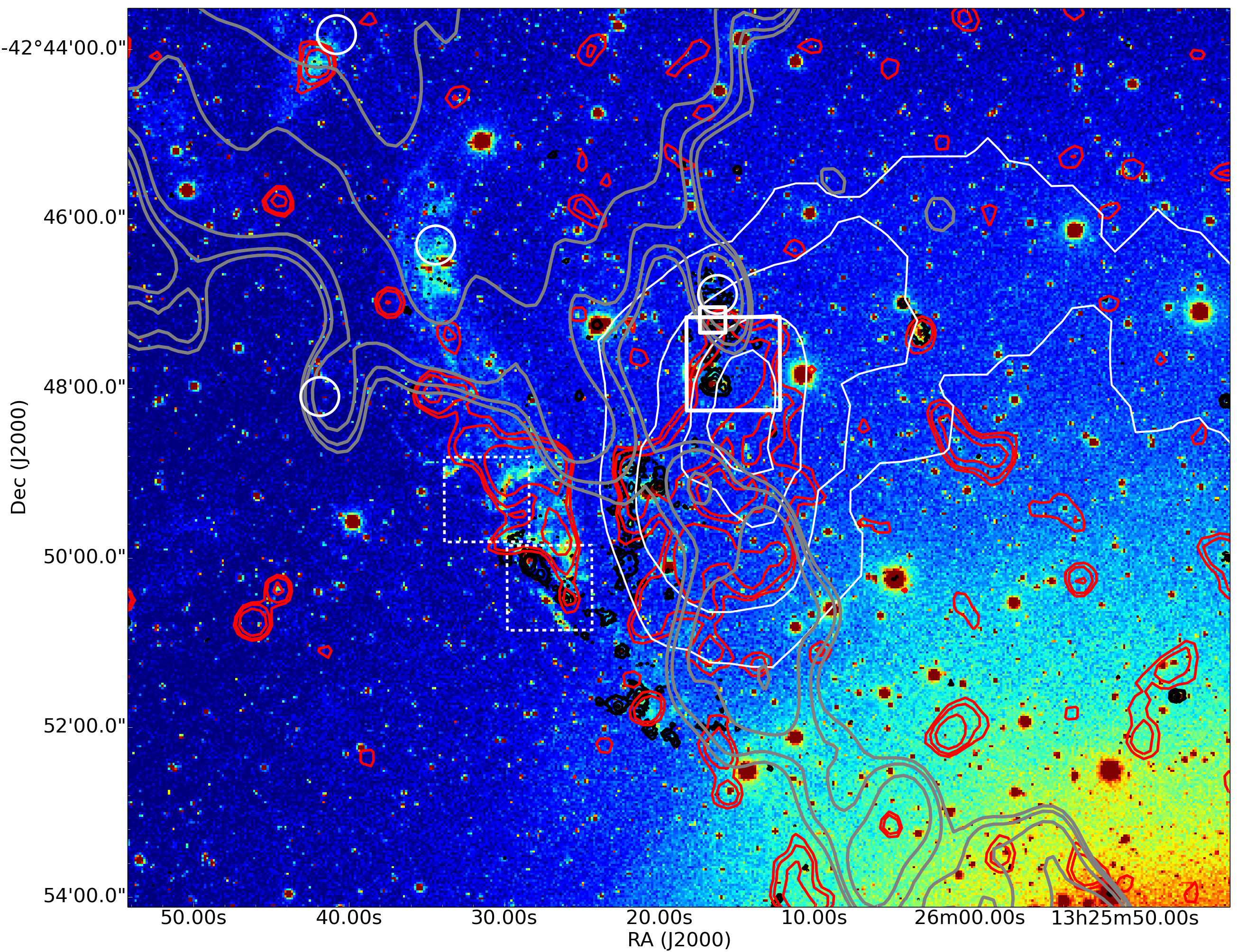}
  \caption{\label{Ha_overview} $H\alpha$ emission of the northern region of Centaurus A with CTIO. We overlaid the H\rmnum{1} emission (VLA; white contours), the $250\: \mu m$ emission (Herschel; red contours), the FUV emission (GALEX; black contours) and the radio continuum (ATCA; grey contours; kindly provided by R. Morganti). The white boxes indicate the location of CO detections with SEST (large box) and ATCA (small box), the dashed boxes are the FOV of MUSE data and the white circles are the positions that were observed with ALMA.}
\end{figure*}

\begin{figure*}[h]
  \centering
  \includegraphics[width=0.44\linewidth]{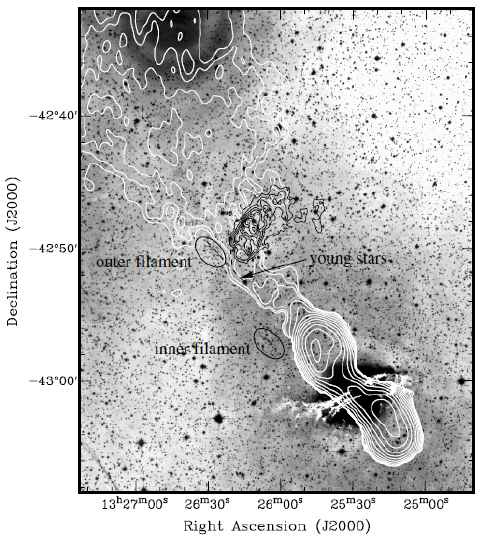}
  \includegraphics[width=0.47\linewidth]{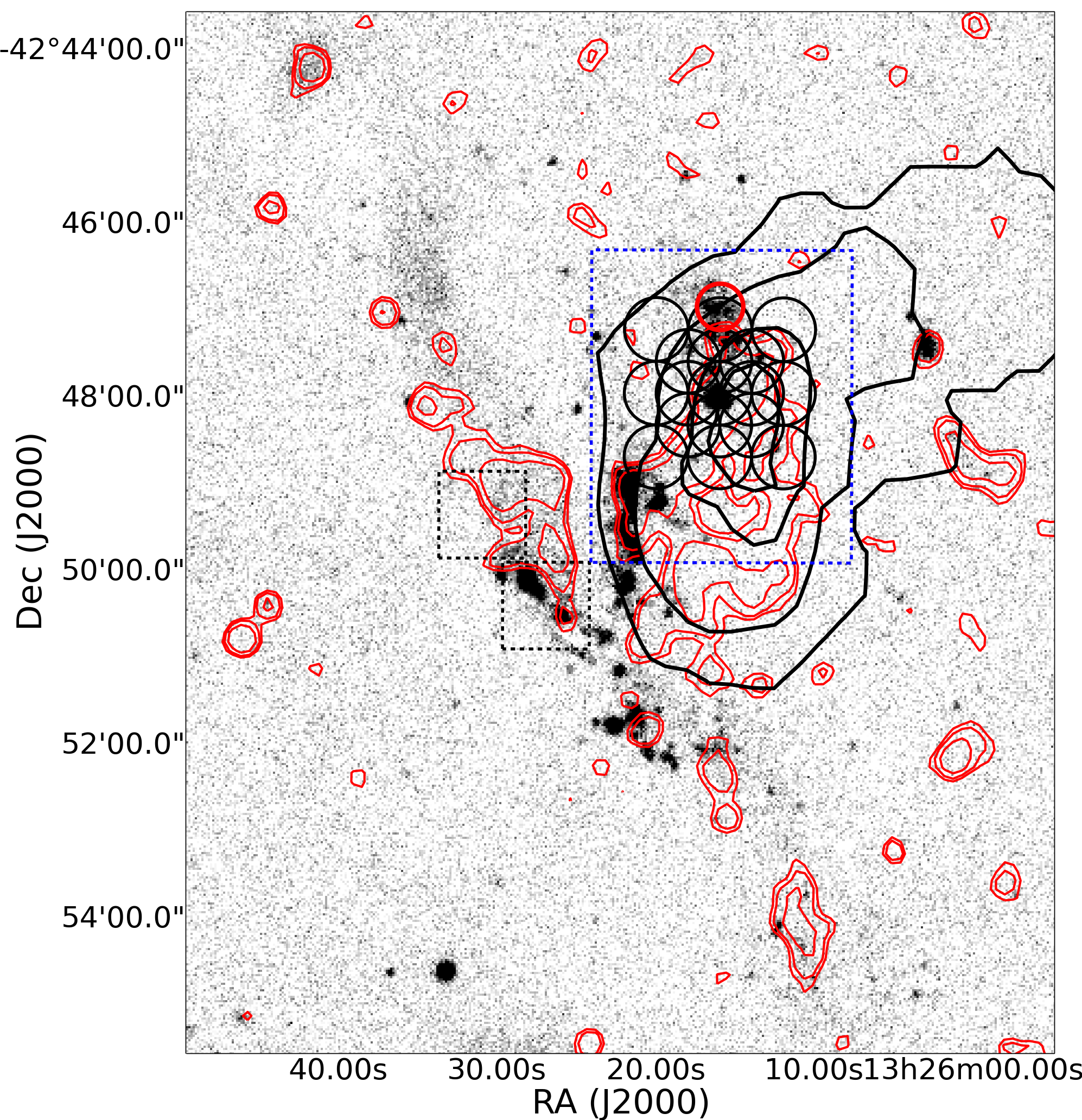}
  \caption{\label{overview} \emph{Left:} Optical image of Cen A in the range $\lambda=3850-5400\: \AA$ (provided by \citealt{Malin_1983} to \citealt{Oosterloo_2005}) showing the diffuse emission and the location of the so-called inner and the outer filaments. The white contours denote the radio continuum emission of the inner lobes and the large-scale jet connecting the northern one to the base of the outer lobe (northern middle lobe). The black contours refer to the north-east outer H\rmnum{1} cloud.
\emph{Right:} FUV image of the outer filament from GALEX. The black and red contours correspond to the H\rmnum{1} and the Herschel-SPIRE $250\: \mu m$ emission, respectively. The dashed boxes show the field of view of MUSE observations \citep{Santoro_2015b}. The blue dashed box indicates the region shown in Fig. \ref{overlay_S1}.}
\end{figure*}

\section{Observations and results}
\label{sec:Obs/res}

   We gathered archival data of the northern filament (Fig. \ref{Ha_overview}) in FUV (GALEX), IR (Herschel) and CO (Swedish-ESO Submillimetre Telescope, SEST, and Atacama Large Millimeter/submillimeter Array ,ALMA). We also searched for HCN and $HCO^+$ (Australian Telescope Compact Array, ATCA) and observed optical emission lines (VLT/MUSE) in different places of the filament. Archival data also include H\rmnum{1} (Very Large Array, VLA) and $H\alpha$ (Cerro Tololo Inter-American Oobservatory, CTIO).

   In Fig. \ref{overlay_S1}, we present the different regions in which CO and HCN/$HCO^+$ were observed on a FUV map. The right panel of Fig. \ref{overview} shows all the outer filaments, where the dust emission at $250\: \mu m$ from Herschel and the H\rmnum{1} shell \citep{Schiminovich_1994} are overlaid to place the observations in a general context.

   The FUV filament is located in the direction of the radio jet. Dust emission is located in projection at the same position of $H\alpha$ and FUV cavities. This gas cloud could be the remnant of a galaxy that merged with Cen A, with pieces still interacting with the radio-jet and wind \citep{Santoro_2015a}.

   \subsection{SFR from TIR and FUV data}

   The area around Centaurus A has been mapped with Herschel \citep{Herschel}. The observation were made with the PACS instrument \citep{PACS} at wavelengths 70 and $160\: \mu m$, and with the SPIRE instrument \citep{SPIRE} at wavelengths 250, 350 and $500\: \mu m$. We used these data, which are available in the online archive (ObsID: 1342188663,1342188855,1342188856). The fluxes of the filament were extracted following the method of \cite{Remy_2013}. The aperture photometry was made in the region of the SEST $44''$ beam; the background was estimated within an annulus of radii $4'-5.5'$ around the filaments. However, a Galactic cirrus strongly influences the background level as the foreground varies in the annulus. Therefore, we extracted the flux for several regions in the annulus and then averaged over the several flux extractions. This leads to uncertainties of $\sim 10\%$.
The filaments present a concentration of young stars (see figure 15 of \citealt{Rejkuba_2001}) that heat the dust. This enable to estimate the SFR along the filaments. If there is a contribution of the old stars in the dust heating, this would only lower the estimated SFR, and also lower the star formation efficiency, thus strengthening our conclusions.

   Cen A was also observed by GALEX \citep{GALEX} in the FUV ($\lambda_{eff}=1539\: \AA, \Delta\lambda=442\: \AA$) and in the near-ultraviolet (NUV: $\lambda_{eff}=2316\: \AA, \Delta\lambda=1060\: \AA$). We only used the FUV here, which we took from the MAST archive. It has a total exposure time of 20101 s. The FUV bandpass is sensitive to O and B stars and is thus a good tracer of star-forming regions. We refer to \cite{Neff_2015b} for a more detailed description of the GALEX data. The FUV flux was converted from count/s to $erg.s^{-1}.cm^{-2}$ by using the formula given on the GALEX website\footnote{http://galexgi.gsfc.nasa.gov/docs/galex/FAQ/counts\_background.html} and the filter bandpass. The fluxes were then corrected for Galactic extinction using the UV extinction determined by \cite{Auld_2012}. We also extracted the flux from the central galaxy and found fluxes consistent with the one derived by \cite{Neff_2015b} in the same region.
\smallskip

   For all the CO positions, we extracted the IR and FUV fluxes. We then used the Herschel data to fit the spectral energy distribution (SED) and determine the IR luminosities. The SED consists of a modified black-body that models dust emission. It was computed with a fixed emissivity index $\beta=1.5$ and gives a dust temperature of $\sim 10-30\: K$.

   The FIR and total infrared (TIR) luminosities were then estimated by integrating the SED over the frequencies between $40-500\: \mu m$ \citep{Garcia_2012} and $3-1100\: \mu m$ \citep{Kennicutt_2012}. The corresponding dust masses leads to molecular gas-to-dust ratios of $\sim 200-600$ in shell S1 (see Table \ref{table:gas-dust}). This wide range of gas-to-dust ratios may be explained by the large uncertainty on the dust mass. The mass is the result of the SED fitting which is sensitive to the uncertainties on the IR background subtraction. We note that the luminosity estimates (and thus the SFR) are less sensitive to flux uncertainties. Using various estimates of the background, we derived uncertainties on the SFR of $\sim 10-30\%$.

   We also derived the FUV luminosity from GALEX. The UV emission emerging from photoionised gas by young and massive stars is often used as a tracer of star formation. The UV luminosity can thus be interpreted as a measure of the star formation rate with $SFR=L_{FUV}/2.24\times 10^{43}\: erg.s^{-1}$ (see \citealt{Kennicutt_2012} for a review).

   The SFR may also be deduced from the TIR luminosity from dust emission. The emission from young stellar population is partly absorbed by dust that heats and emits in TIR through thermal emission. The relation between the TIR luminosity and the SFR is given by $SFR=L_{TIR}/6.7\times 10^9\: L_\odot$ \citep{Kennicutt_2012}. For the IR emission, the SFR can also be derived from the FIR: $SFR=L_{FIR}/5.8\times 10^9\: L_\odot$ \citep{Kennicutt_1998}.

   SFR estimates differ by a factor of up to 10-20, depending on the tracer used. This may be explained by dust absorption that obscures the FUV emission, or it might be that the IR emission is contaminated by Galactic cirrus.

   To solve this problem, we used a multiwavelength estimation of the SFR. We corrected the FUV emission with the TIR emission using a formula from \cite{Kennicutt_2012}: $L_{FUV}^{corr}=L_{FUV}^{obs}+0.46\: L_{TIR}$. The SFR was then calculated from the corrected FUV emission using the conversion formula above.

   \subsection{Filament oxygen abundances as deduced from MUSE data}

\begin{figure*}[htpb]
  \centering
  \includegraphics[width=0.6\linewidth,trim=100 360 120 320,clip=true]{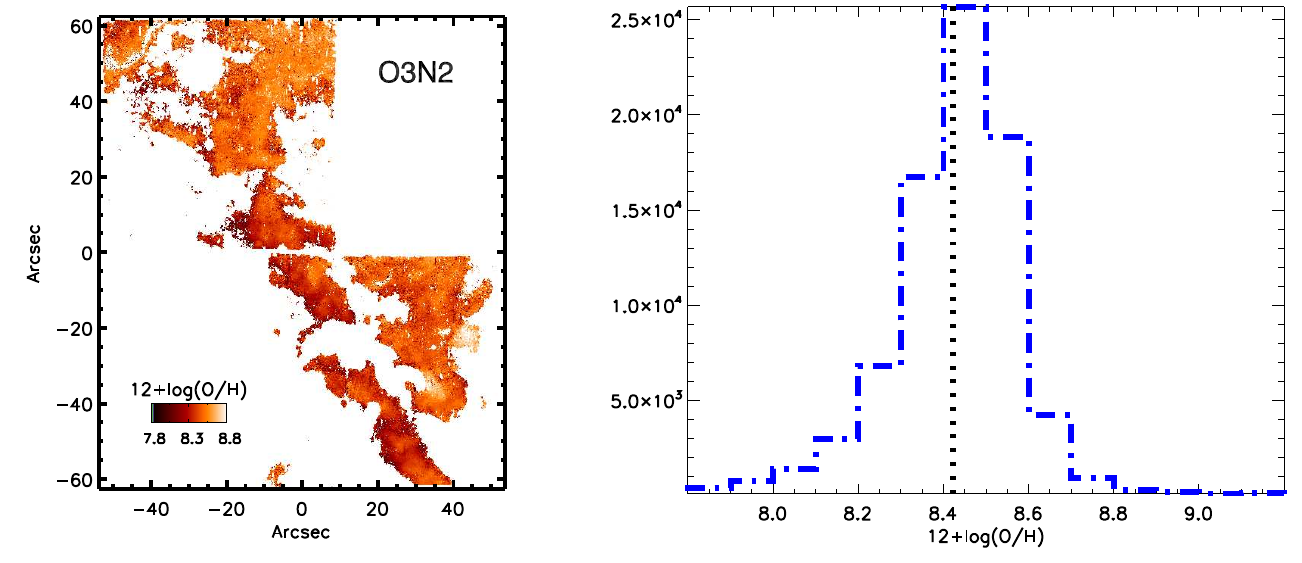} \\
  \includegraphics[width=0.6\linewidth,trim=100 360 120 320,clip=true]{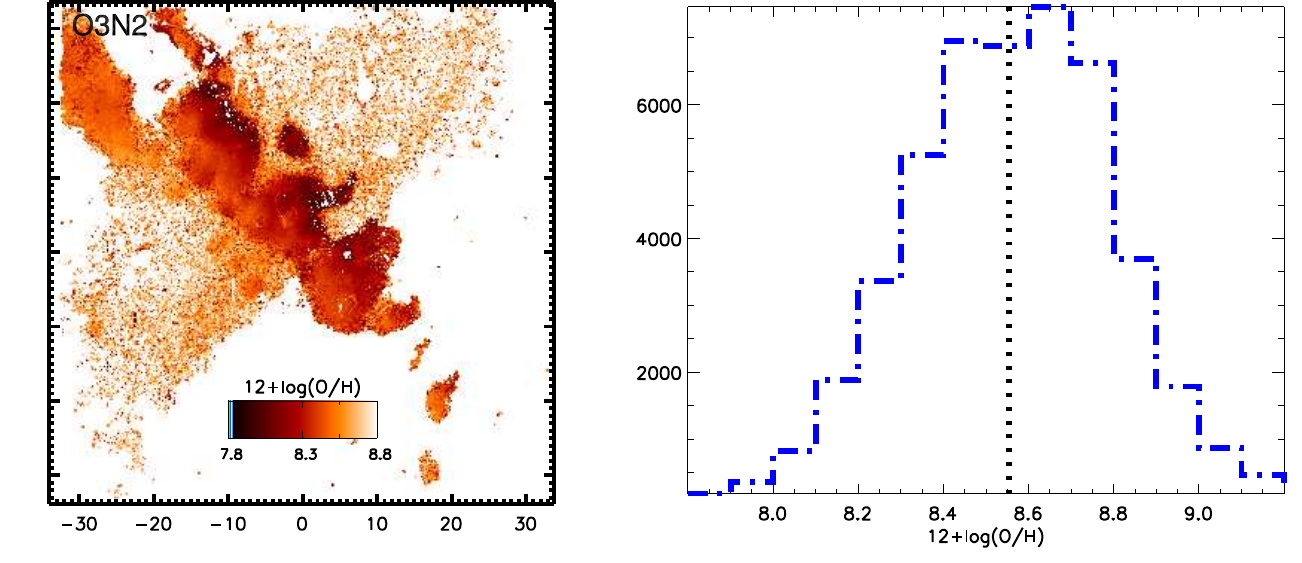}
  \caption{\label{Metallicity} Oxygen abundances derived from [O\rmnum{3}]/[N\rmnum{2}] (left) and histogram of the abundance (right). \emph{Top:} the outer filament region centred at $\alpha=13^h 26^m 29^s.0$, $\delta=-$42:49:51.4; \emph{Bottom:} the inner filament region centred at $\alpha=13^h 26^m 03^s.8$, $\delta=-$42:57:09.7.}
\end{figure*}

   Optical observations have been carried on both the inner and outer filaments with MUSE on the VLT during the science verification period on 25 June 2014 (\citealt{Hamer_2015} for the inner filament, \citealt{Santoro_2015b} for the outer filament). \cite{Hamer_2015} made three pointings with a $3''$ dither and 90\degree rotation between each. \cite{Santoro_2015b} observed two fields in the outer filament of Cen A. The data were reduced with version 0.18.1 of the MUSE data reduction pipeline. Then sky was subtracted using a $20''\times 20''$ region of the field of view (FOV) free from line emission and stars to produce the sky model. Each of the principal lines ($H\alpha$ $\lambda 6562.8$, [N\rmnum{2}] $\lambda 6583$, $H\beta$ $\lambda 4861.3$, [O\rmnum{3}] $\lambda 4959,5007$, [O\rmnum{1}] $\lambda 6366$ and the two [S\rmnum{2}] $\lambda 6716,6731$ lines) was extracted in a separate cube, each covering a velocity range of $\pm 330\: km.s^{-1}$ with a velocity resolution of $ \sim 30\: km.s^{-1}$.
\smallskip

   The flux in each of the principle lines was measured by fitting Gaussian emission line models to the spectra at each spatial resolution element (this procedure is described in detail in \citealt{Hamer_2014}). We calculated oxygen abundances using the $N2$ and $O3N2$ indices following the method of \cite{Pettini_2004}. The oxygen abundance is given by the following equations:
\begin{equation}
  12+\log(O/H)=8.9+0.57\times N2
\end{equation}
and
\begin{equation}
  12+\log(O/H)=8.73-0.32\times O3N2
\end{equation}

\noindent with $N2=\log([N\rmnum{2}]_{\lambda 6583}/H\alpha)$ and $O3N2=\log(\frac{[O\rmnum{3}]_{\lambda 5007}/H\beta)}{([N\rmnum{2}]_{\lambda 6583}/H\alpha})$. Figure \ref{Metallicity} shows the abundances $12+\log(O/H)$ derived using the $O3N2$ factor throughout the inner and outer filaments. These maps show that (i) there is no major metallicity difference between the inner and the outer filaments, (ii) there is no major metallicity gradient along each filament (easier to explain by local or small-scale excitation processes) and (iii) both filaments have relatively high abundances (slightly subsolar) even as far as several kpc away from the centre of NGC 5128.


   In the outer filament, the metallicity seems to be slightly subsolar, with local variations of $12+\log(O/H)$ between 8.3 and 8.6 (Fig. \ref{Metallicity}). \cite{Remy_2014} analysed the dependence of the gas-to-dust (G/D) mass ratio on metallicity. For the outer filament, this would lead to a G/D of $\sim 200-400$. The metallicity-inferred gas-to-dust ratio was used to correct the observed CO luminosity for the CO-dark component \citep{Wolfire_2010,Leroy_2013}.

   \subsection{Single-dish CO observations from the SEST}

   \cite{Charmandaris_2000} made CO observations in May 1999 with the 15m SEST\footnote{The Swedish-ESO Sub-millimetre Telescope was operated jointly by ESO and the Swedish National Facility for Radio Astronomy, Onsala Space Observatory, Chalmers University of Technology} at La Silla. They mapped four regions associated with H\rmnum{1} and stellar shells. One of them covers the intersection of the outer filament with a H\rmnum{1} shell (S1), in a $\sim 2'\times 2'$ map centred at $\alpha=13^h 26^m 16^s.1$, $\delta=-$42:47:55.7. There is a shift of $1'$ in declination between the CO map and the ATCA pointing, that is due to an error in the coordinates given by \cite{Charmandaris_2000}. They observed the CO(1-0) and the CO(2-1) lines, with half-power beam widths of $44''$ and $22''$, respectively. We refer to the article for more details on the observations.

\begin{figure}[h]
  \centering
  \includegraphics[width=\linewidth]{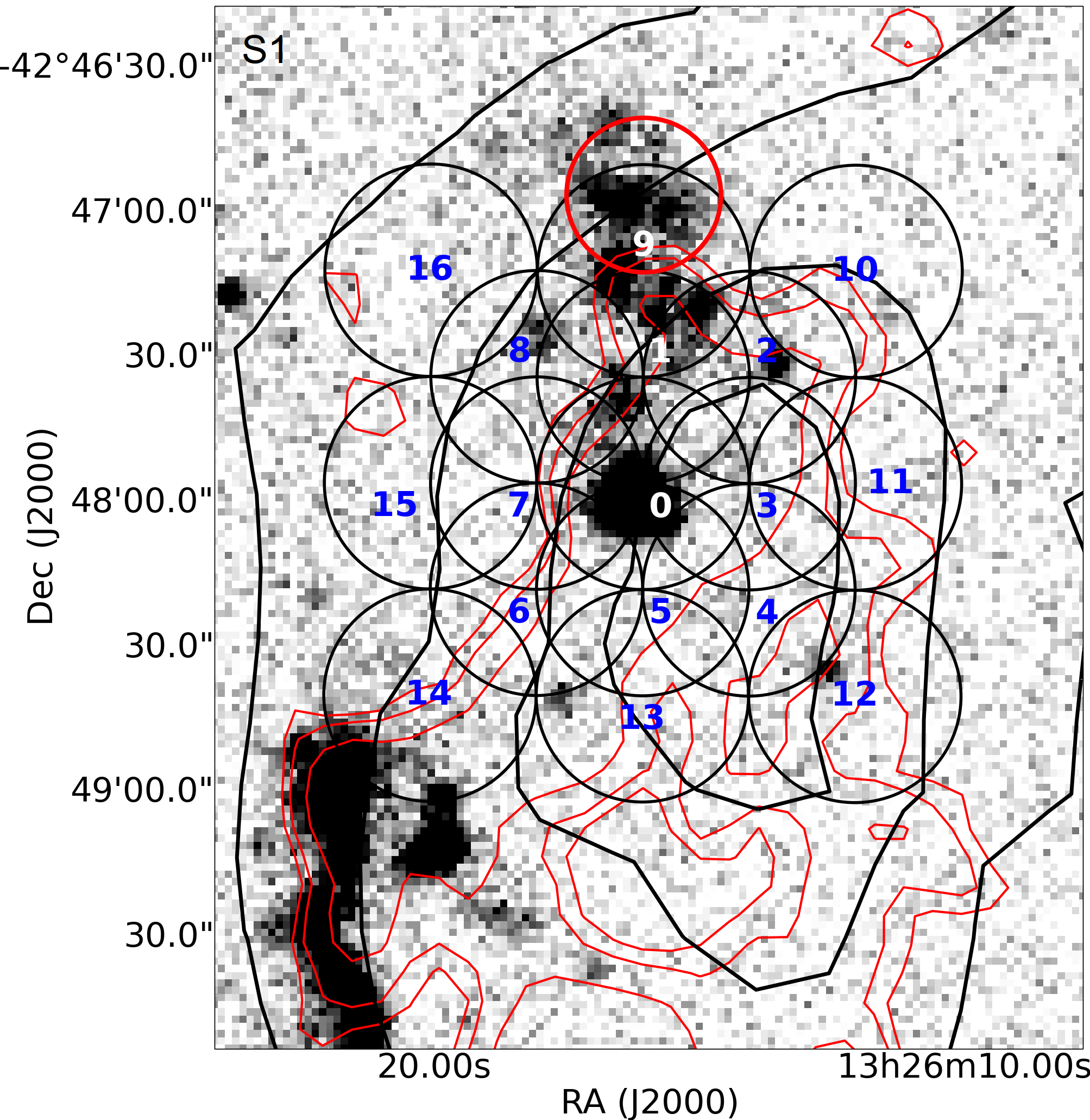}
  \caption{\label{overlay_S1} FUV image from GALEX zoomed on the region observed by \cite{Charmandaris_2000} with SEST (black circles); the red circle indicates the position observed with ATCA. The regions of Table \ref{table:specCO} are labelled by their number.}
\end{figure}

   CO emission was detected in almost all the positions of the $3\times 3$ half-beam central map. Line fluxes were measured by integrating over the channels in the line profile, and the line widths were measured as the full width at 50\% of the peak flux. We estimated the $L'_{CO}$ with the formula from \cite{Solomon_1997}. When there was only a detection in CO(2-1), we assumed a ratio CO(2-1)/CO(1-0) of 0.55 \citep{Charmandaris_2000}.

   CO has also been detected in the two positions $44''$ upward and downward of the central position, including the position overlapping with our ATCA observations. For the other positions that have no detection, upper limits of the line fluxes were calculated at $3\sigma$ with a line width of $18\: km.s^{-1}$ (the frequency resolution). Detailed results for each region are summarised in Table \ref{table:specCO} and compared to TIR and UV data extracted from the same spatial area of diameter $\sim 0.73\: kpc$.

   \subsection{Molecular gas mass}
   \label{sec:CO-to-H2}

   The molecular gas mass was estimated from the line luminosity $L'_{CO}$. A standard Milky Way conversion factor of $4.6\: M_\odot.(K.km.s^{-1}. pc^2)^{-1}$ \citep{Solomon_1997} was used to find molecular gas masses of a few $10^5-10^6\: M_\odot$ in each of the $44''$ ($\sim 0.72\: kpc$) SEST beam. Summing the masses derived in all the positions, we obtain a total molecular gas mass of $\sim 1.4\times 10^7\: M_\odot$, as given in \cite{Charmandaris_2000}.

   However, it is well known that low-metallicity environment may lead to underestimating the amount of molecular gas mass if the standard conversion factor is used \citep{Wolfire_2010}. \cite{Leroy_2013} comprehensively discuss the effect of applying a conversion factor that depends on the D/G on a variety of nearby disk galaxies. The authors showed that applying these corrections leads to more consistent results of the star formation efficiency (inside and among the different galaxies). They also demonstrated that such a metallicity-dependent molecular gas estimation reduces the scatter in the Schmidt-Kennicutt law in nearby disc galaxies. Following \cite{Leroy_2013}, the correction to $\alpha_{CO}$ can be expressed as:
\begin{equation}
  c_{CO-dark}(D/G)=0.65\,exp\frac{0.4}{D/G'\Sigma_{100}}
\end{equation}
with $D/G'$ the dust-to-gas ratio normalised to the Galactic value of 0.01 and $\Sigma_{100}=<\Sigma_{GMC}>/100\: M_\odot.pc^{-2}$. In the following, we assume $\Sigma_{100}=1$ (see discussion in the next section). The MUSE oxygen abundance is subject to local variations between 8.3 and 8.6. These variations lead to a correction of $\alpha_{CO}$ by a factor between 1.4 and 3.2. In the following, we give two mass estimates: one using the standard conversion factor $4.6\: M_\odot.(K.km.s^{-1}. pc^2)^{-1}$ and one using the D/G corrected factor.

   \subsection{Clumpy CO emission highlighted by ALMA}

   We retrieved ALMA data from the science archive : project ADS/JAO.ALMA$\#$2011.0.00454.S. These early science (cycle 0) observations were made in band 6 with 16 antennas and reached a spatial resolution of $2.3''\times 1.3''$, PA -36.8 deg. Six regions of the northern filament were pointed at during $\sim 7.25\: min$ on-source. Only one of these fields coincides with the region S1 studied here. The centre of this ALMA field is the same as for our ATCA observations: $\alpha=13^h 26^m 16^s.1$, $\delta=-$42:46:55.7 (slightly off-centred compared to the brightest CO emission found with the SEST because of the error in the coordinates given in \citealt{Charmandaris_2000}). We used the Common Astronomy Software Applications (CASA) and the supplied script to re-calibrate the data and to produce the final uv-tables in a uvfits format. The data were then transferred into GILDAS to produce datacubes and perform the imaging analysis. The rms reached is $\sim 5\: mJy$ in $\sim 4\: MHz$-channels ($\sim 5\: km.s^{-1}$). The CO(2-1) emission line is detected by ALMA outside the primary beam and reveals three distinct clumps. The size of these clumps is of the order of or smaller than the $<37\times 21\: pc$ synthesised beam resolution. The three clumps show resolved line profiles ($\Delta v\sim 10\: km.s^{-1}$) and are all three dynamically clearly separated by $\sim 10-20\: km.s^{-1}$. Since the emission was found outside the ALMA primary beam (more to the south and closer to the SEST brightest emitting region), all fluxes were scaled with the appropriate primary beam factor ($\times 3.73$). The total integrated flux of the clumps is $S_{CO}\Delta v\sim 3.0\: Jy.km.s^{-1}$. Using the standard $\alpha_{CO}$, this leads to a molecular gas mass of $M_{H_2}\sim 8.7\times 10^4\: M_\odot$. In Table \ref{table:clumps} we summarise the properties of each clump.

\begin{figure}[h]
  \centering
  \includegraphics[width=\linewidth]{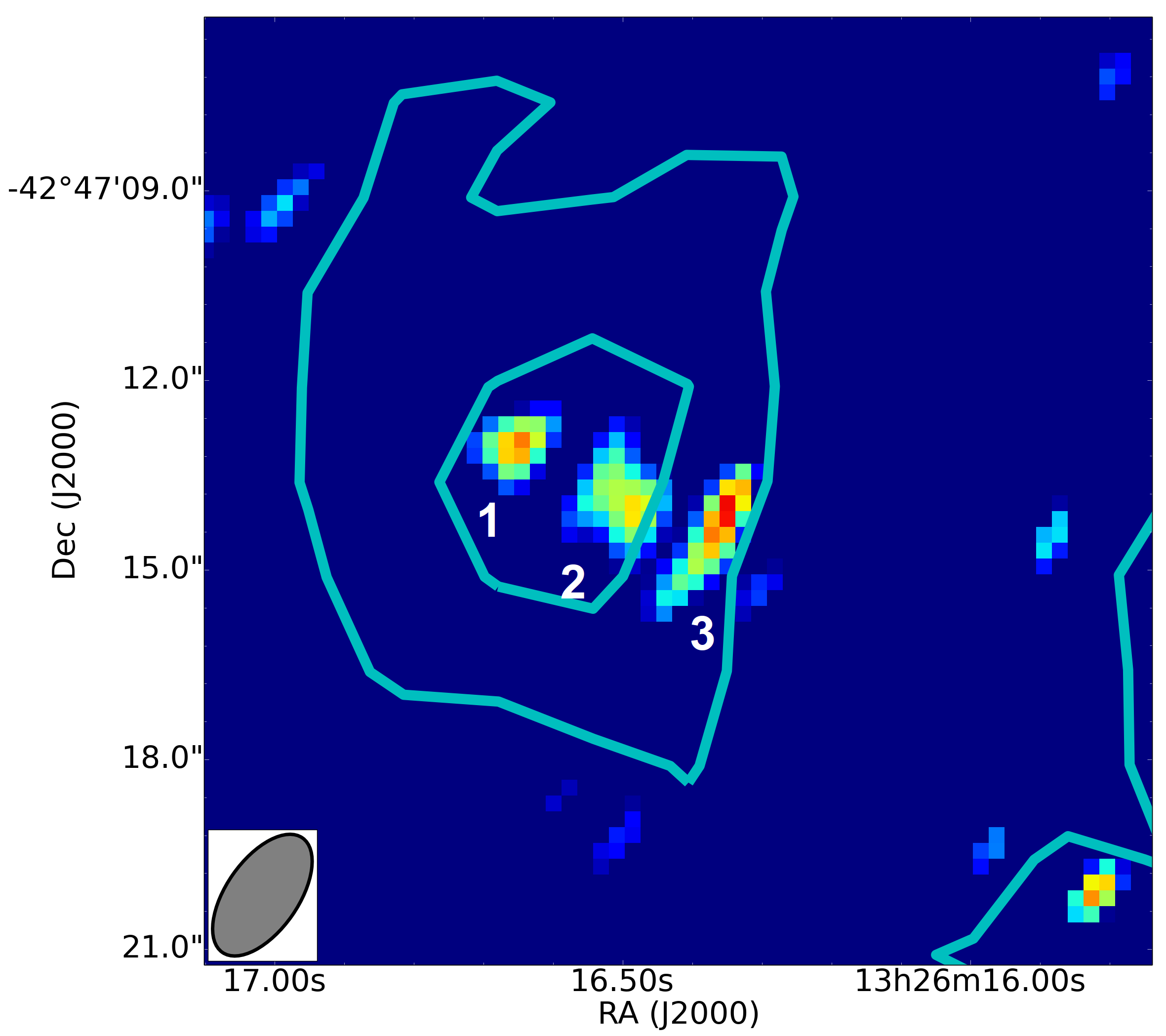}
  \caption{\label{ALMA} ALMA CO(2-1) integrated emission line over $\Delta v\sim 30\: km.s^{-1}$. The contours show the FUV emission from GALEX. The spectra of each clump are shown in Fig. \ref{ALMA-spec}.}
\end{figure}

\begin{figure*}[h]
  \centering 
  \includegraphics[width=0.33\linewidth]{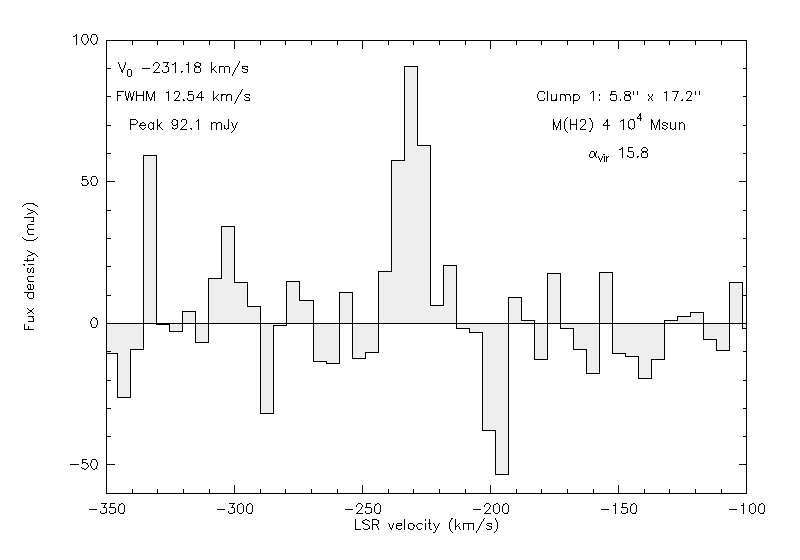}
  \includegraphics[width=0.33\linewidth]{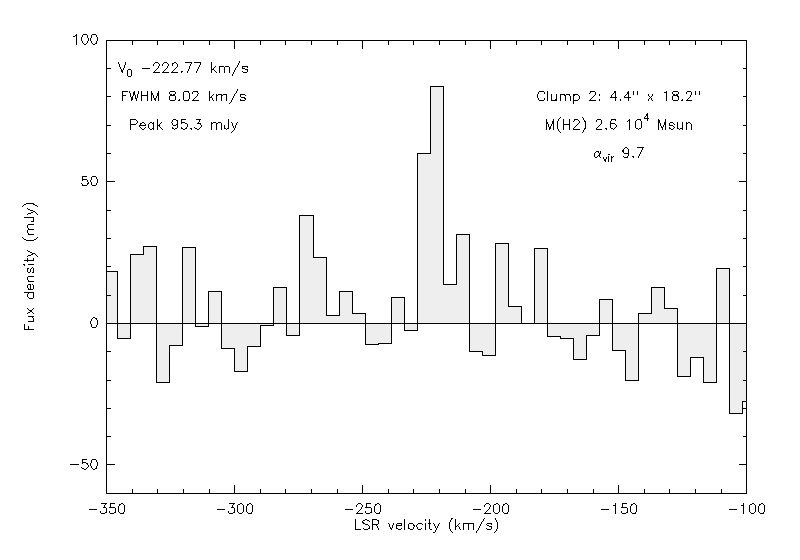}
  \includegraphics[width=0.33\linewidth]{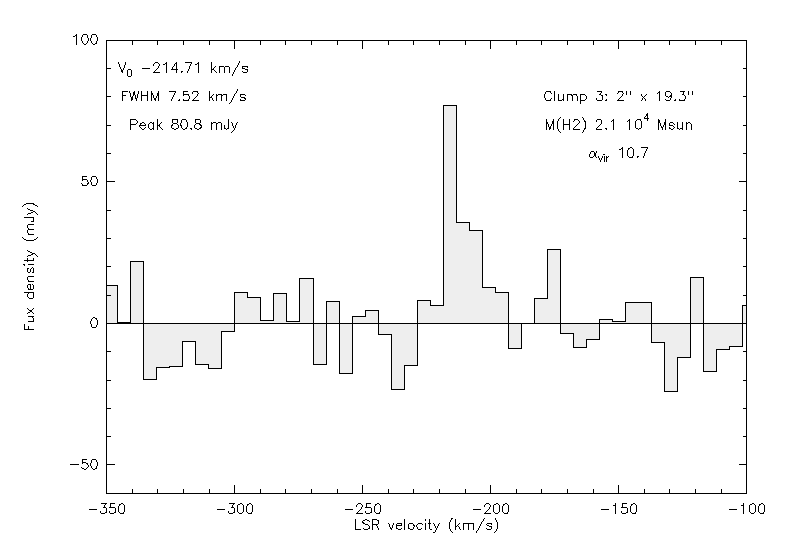}
  \caption{\label{ALMA-spec} ALMA CO(2-1) spectra for each clump numbered in Fig. \ref{ALMA}. The spectral resolution is $\sim 5\: km.s^{-1}$. Characteristics of Gaussian fitting of the line are added on the plot. The rms for each spectrum is $\sim 18\: mJy$ (all fluxes have been corrected for primary beam effect).}
\end{figure*}

\begin{table}[h]
  \centering
  \footnotesize
  \begin{tabular}{lccccc}
    \hline \hline
    \# &      offset      &     $v_0$     &  $\Delta v$   &     $M_{H_2}$      &    $M^Z_{H_2}$     \\
       &                  & ($km.s^{-1}$) & ($km.s^{-1}$) & ($10^4\: M_\odot$) & ($10^4\: M_\odot$) \\ \hline
    1  & $5.8'',-17.25''$ &    $-231$     &     12.5      &    $4.0\pm 1.7$    &    $9.3\pm 3.6$    \\
    2  & $4.4'',-18.2''$  &    $-222$     &      8.0      &    $2.6\pm 1.7$    &    $6.1\pm 2.3$    \\
    3  & $2.0'',-19.3''$  &    $-214$     &      7.5      &    $2.1\pm 1.5$    &    $4.9\pm 1.9$    \\ \hline
  \end{tabular}
  \caption{\label{table:clumps} CO(2-1) emission line properties for each clump (central velocity, FWHM and molecular gas mass estimated with a standard and fixed conversion factor without metallicity correction). Offsets from the ALMA phase centre: $\alpha=13^h 26^m 16^s.1$, $\delta=-$42:46:55.7 are given in the first column.}
\end{table}

   If the CO(2-1) emission is resolved by ALMA, the molecular gas surface density is $\Sigma_{mol}\sim 50-65\: M_\odot.pc^{-2}$. If the emission is not resolved yet (which seems to be the case, at least along the ALMA beam major axis; i.e. giant molecular clouds (GMC) of sizes close to the ALMA minor axis: $\sim 20\: pc$), however, then the surface density could be higher and close to $92-120 M_\odot.pc^{-2}$. Correcting $\alpha_{CO}$ for low metallicity (if necessary) would also lead to an increase of the molecular gas mass and consequently of the surface density. At present we can only conclude that $\Sigma_{mol}>50-65\: M_\odot.pc^{-2}$, which is consistent with common values of nearby galaxies. Using spherical clouds of $\sim 20\: pc$ as a typical radius leads to volume densities of $n_{H_2}\sim 320-420\: cm^{-3}$ (possibly higher if corrections for metallicity are included).

   Taking a Jy/K conversion factor of 7.4 for the ALMA synthesised beam at the observed frequency, we measure a main beam temperature of 0.6 K inside $2.3''\times 1.3''$. For a brightness temperatures of at least 5-10 K, this means a surface filling factor of 0.06-0.12 (a clump radius of 3-5 pc), which is a clumpy distribution with a relatively high density. \\
\cite{Bertoldi_1992} defined a virial parameter $\alpha_{vir}=5\sigma_c^2\, R_c/(GM_c)$ with $\sigma_c$, the velocity dispersion of the cloud, $R_c$, the cloud radius, and $M_c$, its mass. The virial parameter measures the ratio of the kinetic to gravitational energy of the cloud. When $\alpha_{vir}>1$, the clouds are not massive enough to be gravitationally bound. The clumps found in the filaments have $\alpha_{vir}\sim 10-16$ ($8\: km.s^{-1}$, 20 pc, $4\times 10^4\: M_\odot$), which is far from gravitational collapse. The three GMC might be closer to virial equilibrium if a metallicity-corrected $\alpha_{CO}$ were assumed. This would significantly increase the molecular gas mass (of the order of $\sim 2-3$ times).
Evidence of AGN-driven turbulence suppression of star formation has also been found in additional sources \citep{Alatalo_2015a,Guillard_2015}.

   The resolved line widths of three clumps found by ALMA in the northern filament of Cen A show that an input of kinetic energy may have occurred, which makes these clouds relatively inefficient in forming stars despite their high surface density (unless they are at very low metallicity). From the FUV emission, we derive an SFR of $\sim 1.5\times 10^{-5}\: M_\odot.yr^{-1}$, leading to a molecular depletion time of $\sim 5.7\: Gyr$ (and even higher if low metallicity is corrected for). However, below 500 pc, the expected scatter in $t_{dep}$ is very large. The regions of interest are too small to compare the sites of star formation with the reservoir of molecular gas, therefore the estimate of the depletion time for the ALMA clumps should not be over-interpreted. \\
It is not clear whether all the regions inside the filament share the same properties. Five other FOV were covered by the ALMA archived data at the position given in Table \ref{table:ALMA}. All these data have an rms of $5\: mJy$ with a spectral resolution of $5\: km.s^{-1}$. Unfortunately, none corresponds to any bright star-forming region (no star formation tracers are observed, see Fig. \ref{Ha_overview}). High-resolution ALMA mapping of a larger area needs to be performed now to probe possible local variation of the SFE and spatially identify the environmental source of star formation quenching.

\begin{table}[h]
  \centering
  \footnotesize
  \begin{tabular}{ccc}
    \hline \hline
    FOV &      $\alpha$      &   $\delta$    \\ \hline
     1  & $13^h 26^m 41^s.7$ & $-$42:48:06.6 \\
     2  & $13^h 26^m 40^s.5$ & $-$42:43:50.6 \\
     3  & $13^h 26^m 34^s.2$ & $-$42:46:19.8 \\
     4  & $13^h 26^m 25^s.3$ & $-$42:40:17.5 \\
     5  & $13^h 26^m 56^s.8$ & $-$42:41:37.4 \\ \hline
  \end{tabular}
  \caption{\label{table:ALMA} Coordinates of the FOV covered by ALMA in the halo of Centaurus A. Positions 1, 2 and 3 are shown in Fig. \ref{Ha_overview}.}
\end{table}

   Moreover, there is no measurement of the true metallicity at the scale of the ALMA observations. Our MUSE data on several places of the filaments (not coinciding with the ALMA FOV) show that the metallicity is slightly subsolar. If this is the case for the ALMA clumps, then the clouds are not gravitationally bound. Such a turbulent gas inside the galaxy jet or wind might be optically thin \citep{Bolatto_2013}. We therefore computed here the formal lower limit of the molecular gas mass estimated from an optically thin CO emission. Following \cite{Bolatto_2013} for a CO/H$_2$ abundance $Z_{CO}=10^{-4}$ and a $T_{ex}=30\: K$, the CO/H$_2$ conversion factor is $0.34\: M_\odot.(K.km.s^{-1}.pc^2)^{-1}$. The mass of the three clumps would thus be around $10^3\: M_\odot$ and $\alpha_{vir}\sim 150$. It is thus of prime interest to resolve the metallicity map on the ALMA clump to test the turbulent gas scenario.

   \subsection{Dense molecular gas fraction from HCN observations}

   HCN(1-0) and $HCO^+$(1-0), two tracers of the dense molecular gas, have been searched for with the ATCA in June 2014 in the outer filament. Initially, we had planned to point at the intersection with the H\rmnum{1} where CO has been detected, but an error in the coordinates ($\alpha=13^h 26^m 16^s.1$, $\delta=-$42:46:55.7; \citealt{Charmandaris_2000}) induced a shift of $1'$ in declination.

   At redshift z=0.001825, the lines are observable at frequencies of 88.47 GHz and 89.03 GHz, which leads to a primary beam of $32''$. The array was in the H168 configuration (baseline lengths between 61 and 192 m) and the synthesized beam width for the observations at 89 GHz was approximately $3.1''\times 3.1''\sim 50\: pc$. The Compact Array Broadband Backend (CABB) was configured with $2\times 2\: GHz$ bands of 1 MHz resolution, corresponding to a spectral resolution of $3.4\: km.s^{-1}$.

   Data calibration was made with miriad using the standard techniques for ATCA observations. The bandpass calibration was made with PKS 1253-055, while the atmosphere effects were corrected with a nearby phase calibrator (PKS 1424-418). The flux was then scaled with respect to Mars. To improve the signal-to-noise ratio, the data were smoothed to a spectral resolution of $\sim 70\: km.s^{-1}$, and we reached an rms of $\sim 1.0\: mJy$ for an on-source time of $\sim 18h$.
\smallskip

   HCN and $HCO^+$ were not detected with ATCA at the intersection of the outer filament with the H\rmnum{1} shell. We derived upper limits at $3\sigma$ of $\sim 3\: mJy$ at both 88.5 and 89 GHz in the $3.1''$ synthesised beam. Assuming a line width of the order of the one detected in CO ($\sim 10\: km.s^{-1}$), the integrated flux are $S_{HCN}\Delta v<33.5\: mJy.km.s^{-1}$ and $S_{HCO^+}\Delta v<33\: mJy.km.s^{-1}$ in the synthesised beam of $3.1''$. Using the formula from \cite{Solomon_1997}, we calculated upper limits of the luminosities. The results are summarised in Table \ref{table:specHCN} and give $L'_{HCN}<1.6\times 10^3\: K.km.s^{-1}.pc^2$ and $L'_{HCO^+}<1.6\times 10^3\: K.km.s^{-1}.pc^2$.

   The dense molecular gas mass was estimated from the line luminosity $L'_{HCN}$. The conversion factor is poorly constrained as there is no direct calibration from giant molecular clouds. Using the large velocity gradient and virial calculations, \cite{Gao_2004a} derived a conversion factor of $10\: M_\odot.(K.km.s^{-1}. pc^2)^{-1}$ for a Milky Way-like galaxy. Using the same conversion factor, we found an upper limit of the dense molecular gas mass of a few $10^4\: M_\odot$.

\begin{table}[h]
  \centering
  \footnotesize
  \begin{tabular}{lcccc}
    \hline \hline
    Region &       Area       &      $L'_{CO}$       &      $L'_{HCN}$      & $\cbra{\frac{L'_{HCN}}{L'_{CO}}}$ \\
           &     ($pc^2$)     & ($K.km.s^{-1}.pc^2$) & ($K.km.s^{-1}.pc^2$) &                                   \\ \hline
    Cen A  & $3.1\times 10^8$ &   $7.2\times 10^7$   &   $5.5\times 10^6$   &            $\sim 0.08$            \\
    Outer  & $8.2\times 10^3$ &   $2.1\times 10^4$   &  $<4.8\times 10^3$   &              $<0.23$              \\ \hline
  \end{tabular}
  \caption{\label{table:specHCN} CO and HCN luminosities in the central galaxy \citep{Wild_2000,Espada_2009} and in the region of the ALMA/ATCA observations. The $L'_{CO}$ from ALMA corresponds to the three clumps. For HCN with ATCA, an upper limit is computed at $3\sigma$ in three $3.1''$ ATCA synthesised beam, with a line width of $\Delta v=10\: km.s^{-1}$.}
\end{table}

   We define the dense gas-depletion time as the time needed to consume all the dense gas with the present star formation rate: $t_{dense}\sim M_{dense}/SFR$. As the dense gas mass at the ATCA pointing is $<4.8\times 10^4\: M_\odot$, the depletion time of the dense gas is $<3.2\times 10^9\: yr$. This upper limit on the depletion time is about 50 times higher than what was found for star-forming spiral galaxies and about 175 times higher than for luminous and ultraluminous infrared galaxies (LIRG/ULIRG) \citep{Garcia_2012}. This means that the present data do not allow constraining the star formation efficiency from the dense gas in a relevant way. There is no sign of a particularly increased star formation efficiency in the observed region, however. Deeper observations with ALMA are required to obtain a higher sensitivity and cover a larger region.
\smallskip

   We also averaged the ATCA data over the whole bandwidth ($\sim 2\: GHz$) to derive a continuum map in the millimetre domain. No sign of continuum emission from the filament is found in this map. The rms of the ATCA continuum map is $\sim 0.15\: mJy/beam$ at 87.8 GHz and $\sim 0.1\: mJy/beam$ at 89.8 GHz, in a beam of $3.1''$. With the VLA, \cite{Neff_2015b} found in the same region a continuum emission of $\sim 200\: mJy/beam$ at 327 MHz in a beam of $28''\times 17''$, which must be dominated by the non-thermal synchrotron. In the mm-domain, the emission is more likely dominated by a mixture of free-free and thermal dust emission. Following \cite{Murphy_2011}, we used the 3mm continuum upper limit to estimate an upper limit on the $SFR^T$ as measured from the free-free emission. We assumed $T_e=10^4\: K$ and $L_v^T<8\times 10^{23}\: erg.s^{-1}.Hz^{-1}$, with the free-free flux assumed to be $\sim 60\%$ of the continuum upper limit at 89.8 GHz. This leads to an $SFR^T<6\times 10^{-4}\: M_\odot.yr^{-1}$, a value consistent with the SFR measured in region 9 from the UV and the TIR. Mapping the free-free emission of these regions with such a low star formation rate is beyond the reach of current facilities and would remain difficult even with ALMA when fully operational.

\section{Discussion}
\label{sec:discussion}

   \subsection{Kennicutt-Schmidt law}
   \label{sec:SFE}

\begin{figure*}[h]
  \centering
  \includegraphics[page=1,width=0.49\linewidth,trim=25 0 45 25,clip=true]{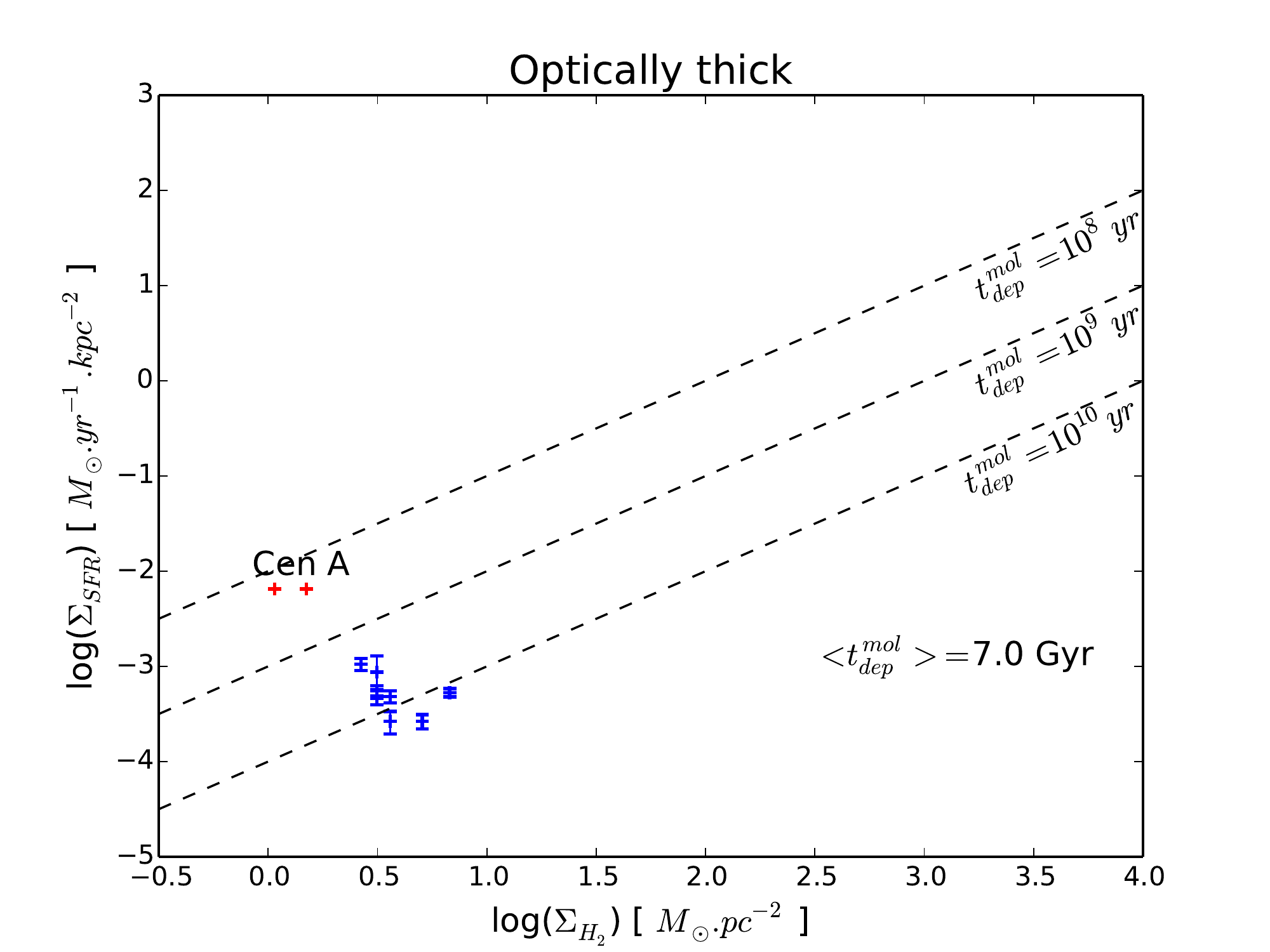}
  \includegraphics[page=3,width=0.49\linewidth,trim=25 0 45 25,clip=true]{SK-law.pdf}
  \caption{\label{KS-law} $\Sigma_{SFR}$ vs. $\Sigma_{gas}$ for the different regions of CO emission. The diagonal dashed lines show lines of constant SFE, indicating the level of $\Sigma_{SFR}$ needed to consume 1\%, 10\%, and 100\% of the gas reservoir in $10^8$ years. Thus, the lines also correspond to constant gas depletion times of, from top to bottom, $10^8$, $10^9$, and $10^{10}\: yr$. The red crosses correspond to the central galaxy. Two cases have been studied: (1) optically thick CO emission with standard $\alpha_{CO}$ (\emph{left}), and (2) optically thick CO emission corrected for metallicity (\emph{right}). To avoid beam effects, we only used the CO(1-0) data (without regions 8 and 9).}
\end{figure*}

   We calculated the molecular gas and SFR surface densities ($\Sigma_{H_2}$, $\Sigma_{SFR}$) for all regions mentioned in \cite{Charmandaris_2000}. The two quantities were smoothed over the SEST $44''$ beam, except for regions 8 and 9, where the mass density was smoothed over the SEST $22''$ beam. We then plotted the $\Sigma_{SFR}$ vs $\Sigma_{H_2}$ diagram (see Fig. \ref{KS-law}, \citealt{Bigiel_2008,Daddi_2010}) for the standard and metallicity corrected $\alpha_{CO}$. For comparison, we also plot the central galaxy in Fig. \ref{KS-law}. \cite{Espada_2009} found a gas surface density of $1.5\: M_\odot.pc^{-2}$ in a region of $\sim 10\: kpc$ radius. Using the same region with the mass derived by \cite{Eckart_1990}, we found a surface density $\Sigma_{H_2}\sim 1.07\: M_\odot.pc^{-2}$. \cite{Neff_2015b} derived an SFR of $\sim 2\: M_\odot.yr^{-1}$, leading to a surface density $\Sigma_{SFR}\sim 6.5\times 10^{-3}\: M_\odot.yr^{-1}.kpc^{-2}$.

   First, we see that the central galaxy is forming stars very efficiently, similar to ULIRG. For the positions in the filaments where CO has been detected, the regions seem to follow a Schmidt-Kennicutt law $\Sigma_{SFR}\propto \Sigma_{H_2}^N$ \citep{Kennicutt_1998} regardless of the $\alpha_{CO}$, which lies lower than star-forming spiral galaxies, with depletion times of 7.0 Gyr and 16.2 Gyr on average, for a standard and metallicity-corrected $\alpha_{CO}$, respectively.

   \subsection{Molecular gas depletion time}

   The depletion time is the time in which all the molecular gas would be consumed if the SFR remained constant: $t_{dep}^{mol}\sim M_{H_2}/SFR$. When using a fixed standard $\alpha_{CO}$, we find an average depletion time of 7.0 Gyr for all SEST positions. This depletion time is much longer when a metallicity correction of the conversion factor is included. We find an average of 16.2 Gyr. The two last columns of Table \ref{table:specCO} shows the SFR and depletion time at the different positions. In the Milky Way and in nearby galaxies, scaling relations between SFR and $H_2$ (e.g. Kennicutt-Schmidt law) are usually averaged relations because many distinct regions are mixed. \cite{Leroy_2013} studied the scale effects by measuring the dependence of the scatter of $t_{dep}$ on the observed spatial resolution. They discussed that the expected rms scatter of $\log t_{dep}$ follows the relation $\sigma(l)=\sigma_{600}\cbra{\frac{l}{600\: pc}}^{-\beta}$, where l represents the spatial resolution and $\sigma_{600}$ is the scatter in $t_{dep}$ at 600 pc resolution. The power-law index $\beta$ measures the rate at which changing the resolution changes the measured scatter. For uncorrelated star formation, $\beta=1$ is expected. Any deviation from this value measures the effect of spatially correlated depletion time caused by an environmental or a morphological phenomenon that would synchronize the star formation.

   To track scale effects, we estimated the scatter of $\log t_{dep}^{mol}$ using a similar method. As we do not have a map of CO emission, we could not change the spatial resolution. We therefore estimated the scatter at two scales by combining the pointings of our map: (1) for a single beam and (2) for a combination of $2\times 2$ beams. The depletion time scatter is consistent with $\beta=1$, which is an indication of spatially uncorrelated star formation. This means that there is no indication of any morphological effect on the star formation efficiency along the SEST pointings.

   \subsection{Jet-cloud interaction}

\begin{figure}[h]
  \centering
  \includegraphics[width=0.8\linewidth]{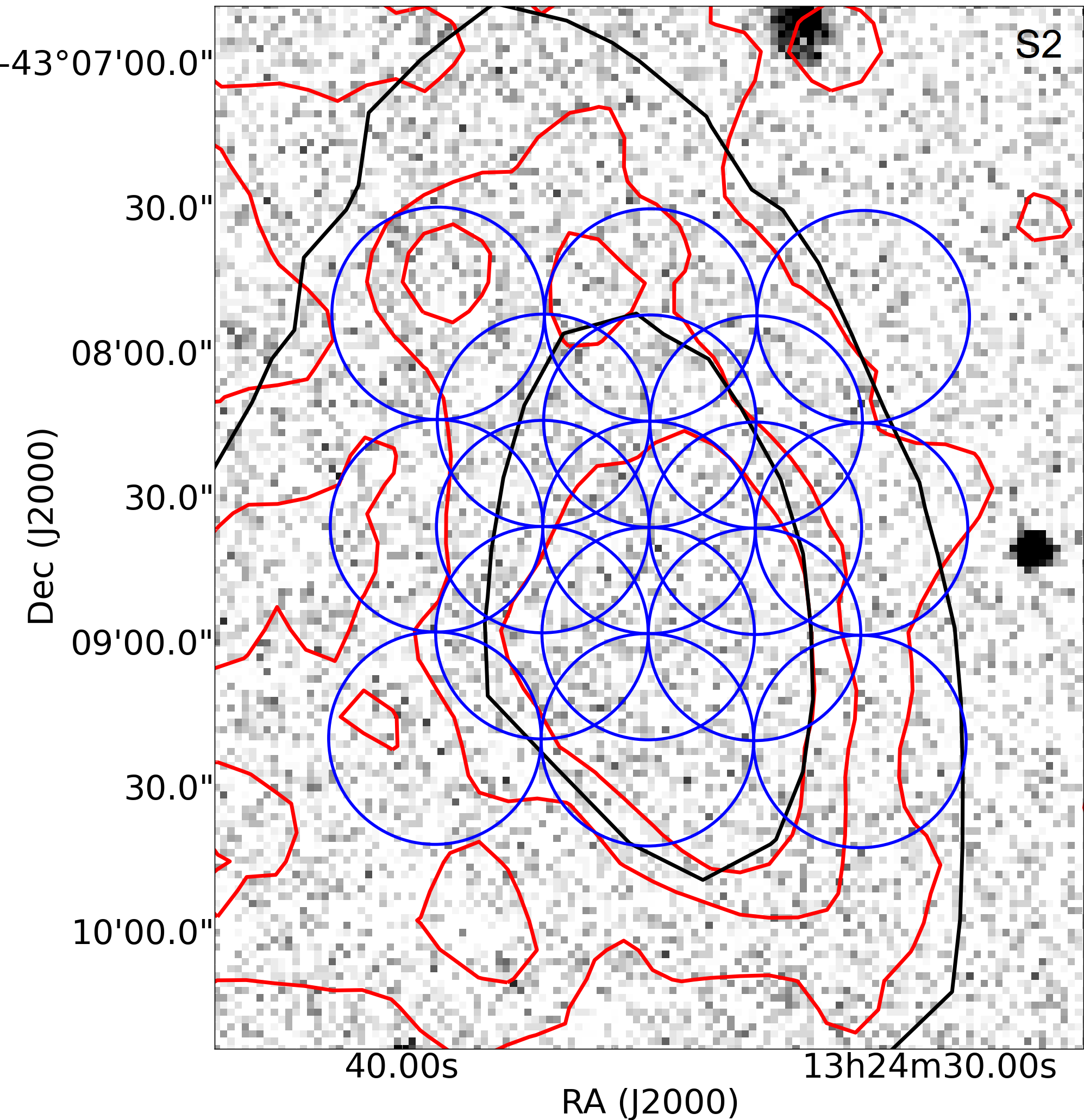}
  \caption{\label{overlay_S2} FUV image from GALEX of the region around shell S2 in the south. The circles show the position observed with SEST (blue: \cite{Charmandaris_2000}). The black and red contours correspond to the H\rmnum{1} and the Herschel-SPIRE $250\: \mu m$ emission, respectively.}
\end{figure}

   To find more evidence of jet-cloud interaction in the outer filaments, we conducted the same study in the southern shell (S2 of \cite{Charmandaris_2000}, see Fig. \ref{overlay_S2}). This region shows no FUV emission which could be fully attenuated by the foreground dust (cirrus). However, \cite{Auld_2012} derived the FUV extinction in the two shells S1 and S2 and found a small difference. This indicates that if the FUV emission were as bright as in S1, it would not be fully attenuated by Galactic cirrus. Therefore, the SFR would be undetectable and lower than in S1. \\
Using the TIR emission as a star formation tracer, we derived an SFR similar to or higher in S2 than in S1. However, this part of the sky is contaminated by cirrus emission, which prevents any robust estimation of the TIR emission from the shell. The dust peak found with Herschel nevertheless overlays the region found in CO with the SEST. This strongly suggests that star formation occurs in the southern filament, along the jet direction, at the place of some molecular gas. It is qualitatively consistent with our conclusion for the northern filament even if we cannot quantitatively and accurately determine the local morphology of the efficient star-forming regions.
Table \ref{table:CO_S2} summarises the masses derived from CO and the SFR derived from the TIR emission; but these results are subject to high uncertainties.

   The third region detected in HI (S3 of \cite{Charmandaris_2000}) is not aligned with the jet direction. This region is not detected in CO by the SEST and there is no evidence of any star formation activity either in the GALEX or the Herschel data. This latest result suggests that the jet-gas interaction may indeed be necessary to compress and cool the gas or dissipate the local turbulence so that stars may be formed. Detailed and deeper observations of the northern shell are being investigated to answer this question.

   Figure \ref{Ha_overview} summarises the different tracers of star formation in the outer filament. The western part presents FUV emission, the eastern part does not, but emits $H\alpha$ emission.

   \subsection{Origin of the filaments}

   \cite{Neff_2015b} summarised and discussed the different possible origins of the filaments in Centaurus A. The favoured scenario proposes that the filaments are the radiating interface of a remnant companion, disrupted during a recent collision with NGC 5128. The material falling onto the central galaxy could indeed be ionised and shocked when crossing the AGN-jet or stellar-wind-exciting cone. The relatively widely spread metallicity of the filaments measured by MUSE supports the scenario of some enriched material, either coming from a companion galaxy or lifted up from NGC 5128 itself. The metallicity of both filaments can also be explained if they are made of pieces of the same companion. The H\rmnum{1} cloudlets found in absorption by \cite{Struve_2010} inside the jet cone and closer to the central galaxy are another evidence of the presence of cold gas at different places along the jet-wind interaction cone. The dust lane and the relatively large amount of molecular gas in NGC 5128 is also interpreted as due to the recent cannibalism of a small gas-rich companion. Determining the origin of the filaments and of the molecular gas found at distances as far as 15 kpc from the central galaxy is beyond the scope of the present discussion. In the following, we use the molecular gas mass and the star formation tracers along the radio-jet as a testbed of the effect of the jet-wind interaction on the cold gas outside the galaxy (triggering or quenching star formation).

\section{Conclusion}

   The star formation rate in the northern filament is only $4\times 10^{-3}\: M_\odot.yr^{-1}$ in the whole region. This is much lower than in the centre of Centaurus A ($\sim 2\: M_\odot.yr^{-1}$), that is, around $\sim 0.2\%$. This means that even if the radio-jet could trigger star formation outside the galaxy, the effect would appear negligible at the galaxy scale. The mass in the whole filament is $>1.7\times 10^7\: M_\odot$, that is $\sim 5\%$ of total gas mass in Centaurus A ($3.3\times 10^8\: M_\odot$; \citealt{Charmandaris_2000}). On a large scale, the star formation efficiency is therefore higher in the central galaxy (most likely because of the recent merger) than in the filaments.

   We here carried out a detailed analysis of archival GALEX and Herschel data to derive star formation proxies at a sub-kpc scale inside the northern filament in Centaurus A. Our goal was to compare the local star formation rate to the amount of molecular gas detected in the same region, which was not described in detail by \cite{Charmandaris_2000}. We also retrieved ALMA observations of the CO(2-1) line of the same region. To derive the molecular gas masses, we used a standard conversion factor and a metallicity-corrected conversion factor based on the dust-to-gas ratio. The metallicity variations in the filaments (not overlapping fields) were estimated with MUSE data, assuming the whole filaments have the same metallicity (slightly subsolar; $12+\log(O/H)=8.3-8.6$). We found that the gas and SFR surface densities follow a KS-relation with long depletion times: 7 Gyr and 16 Gyr on average (without and with metallicity-corrected $\alpha_{CO}$).

   We performed ATCA observations to detect dense gas tracer emission (HCN, HCO+), but we were unable to detect any sign of star formation efficiency enhancement in the dense gas.

   High-resolution ALMA observations of the CO(2-1) emission revealed three unresolved clumps of sizes $<37\times 21\: pc$, with a line width of the order of $10 km.s^{-1}$. (1) The mass of these clumps was evaluated to be $\sim 10^4\: M_\odot$ with a standard conversion factor. This leads to high values of the virial parameter \citep{Bertoldi_1992}, tracing a turbulent gas. The AGN jet or wind may be the cause of this input of kinetic energy, leading to very inefficient star-forming molecular gas regions. (2) However the scatter in gas mass estimates due to low-metallicity correction could be enough to increase the real mass of gas and lead to gravitationally bound clouds. (3) The last and more formal hypothesis is that this region is indeed of solar metallicty, so that the turbulent increase of line width is really large compared to the mass. In this case, an optically thin estimate of the gas mass would even increase the effect: lowering the molecular gas mass would lead to very large virial parameter together with a very efficient star formation (short depletion time), which seems difficult to reconcile.

   We finally looked for the morphological structure of the star formation efficiency (with the CO line ratio and gas surface density), but we were enable to find any reliable trend. The jet-cloud interaction seems to quench star formation in the molecular gas lying along the jet or wind trajectory, at least in the regions observed. Whether this is the case everywhere in the filament is still an open question. A better coverage of the filaments at high resolution, combining molecular gas mass probes (CO measurements and metallicity tracers) is now necessary to confirm whether this trend applies to the whole filament or if some regions can still form stars efficiently. Another paper based on APEX data of the whole filament is currently in preparation.


\begin{table*}[h]
  \centering
  \begin{tabular}{llccccc|cc}
    \hline \hline
\multicolumn{2}{l}{Position} &     $I_{CO}$    &      $L'_{CO}$       &     $M_{H_2}$     &        $SFR$        & $t_{dep}^{mol}$ & $M^Z_{H_2}$  & $t^Z_{dep}$ \\
\multicolumn{2}{l}{(offset)} & ($K.km.s^{-1}$) & ($K.km.s^{-1}.pc^2$) &    ($M_\odot$)    & ($M_\odot.yr^{-1}$) &      (Gyr)      & ($M_\odot$)  &    (Gyr)    \\ \hline
    0          & ($0'',0''$)     &      1.14       &   $6.1\times 10^5$   & $2.8\times 10^6$  & $2.2\times 10^{-4}$  &    12.7   & $6.5\times 10^6$  &   29.7  \\
    1          & ($0'',22''$)    &      0.53       &   $2.8\times 10^5$   & $1.3\times 10^6$  & $2.0\times 10^{-4}$  &    6.50   & $3.0\times 10^6$  &   15.1  \\
    2          & ($-22'',22''$)  &      0.53       &   $2.8\times 10^5$   & $1.3\times 10^6$  & $2.3\times 10^{-4}$  &    5.65   & $3.0\times 10^6$  &   13.2  \\
    3          & ($-22'',0''$)   &      0.46       &   $2.4\times 10^5$   & $1.1\times 10^6$  & $4.4\times 10^{-4}$  &    2.50   & $2.6\times 10^6$  &   5.83  \\
    4          & ($-22'',-22''$) &      0.53       &   $2.8\times 10^5$   & $1.3\times 10^6$  & $3.6\times 10^{-4}$  &    3.61   & $3.0\times 10^6$  &   8.41  \\
    5          & ($0'',-22''$)   &      0.61       &   $3.2\times 10^5$   & $1.5\times 10^6$  & $2.0\times 10^{-4}$  &    7.50   & $3.5\times 10^6$  &   17.5  \\
    6          & ($22'',-22''$)  &      0.86       &   $4.5\times 10^5$   & $2.1\times 10^6$  & $1.1\times 10^{-4}$  &    19.1   & $4.9\times 10^6$  &   44.5  \\
    7          & ($22'',0''$)    &     $<0.19$     &  $<1.0\times 10^5$   & $<4.6\times 10^5$ & $7.2\times 10^{-5}$  &  $<6.39$  & $<1.1\times 10^6$ & $<14.9$ \\
    8          & ($22'',22''$)   &     $<0.19$     &  $<1.0\times 10^5$   & $2.5\times 10^5$  & $3.2\times 10^{-5}$  &    10.9   & $8.2\times 10^5$  &   25.5  \\
    9          & ($0'',44''$)    &     $<0.19$     &  $<1.0\times 10^5$   & $4.4\times 10^5$  & $1.0\times 10^{-4}$  &    4.40   & $1.0\times 10^6$  &   10.3  \\
    10         & ($-44'',44''$)  &     $<0.19$     &  $<1.0\times 10^5$   & $<4.6\times 10^5$ & $<7.0\times 10^{-6}$ &     -     & $<1.1\times 10^6$ &    -    \\
    11         & ($-44'',0''$)   &     $<0.19$     &  $<1.0\times 10^5$   & $<4.6\times 10^5$ & $5.5\times 10^{-5}$  &  $<8.36$  & $<1.1\times 10^6$ & $<19.5$ \\
    12         & ($-44'',-44''$) &     $<0.19$     &  $<1.0\times 10^5$   & $<4.6\times 10^5$ & $1.4\times 10^{-4}$  &  $<3.29$  & $<1.1\times 10^6$ & $<7.66$ \\
    13         & ($0'',-44''$)   &      0.61       &   $3.2\times 10^5$   & $1.5\times 10^6$  & $1.1\times 10^{-4}$  &    13.6   & $3.5\times 10^6$  &   31.8  \\
    14         & ($44'',-44''$)  &     $<0.19$     &  $<1.0\times 10^5$   & $<4.6\times 10^5$ & $1.1\times 10^{-4}$  &  $<4.18$  & $<1.1\times 10^6$ & $<9.74$ \\
    15         & ($44'',0''$)    &     $<0.19$     &  $<1.0\times 10^5$   & $<4.6\times 10^5$ & $<7.5\times 10^{-6}$ &     -     & $<1.1\times 10^6$ &    -    \\
    16         & ($44'',44''$)   &     $<0.19$     &  $<1.0\times 10^5$   & $<4.6\times 10^5$ & $<8.5\times 10^{-6}$ &     -     & $<1.1\times 10^6$ &    -    \\ \hline
  \end{tabular}
  \caption{\label{table:specCO} CO luminosities and molecular masses in the regions observed in shell S1 by \cite{Charmandaris_2000} that show evidence of star formation, either through TIR or FUV emission. $M_{H_2}$ is the molecular mass derived from the CO(1-0) emission, except for regions 8 and 9 (CO(2-1)). The SFR and $t_{dep}^{mol}$ were estimated in regions of $44''$, corresponding to the beam of SEST for CO(1-0). The offset ($0'',0''$) is centred on $\alpha=13^h 26^m 16^s.1$, $\delta=-$42:47:55.7.}
\end{table*}

\begin{table*}[h]
  \centering
  \begin{tabular}{llccccc}
    \hline \hline
    \multicolumn{2}{l}{Position} &     $I_{CO}$    &      $L'_{CO}$       &     $M_{H_2}$     &        $SFR$         & $t_{dep}^{mol}$ \\
    \multicolumn{2}{l}{(offset)} & ($K.km.s^{-1}$) & ($K.km.s^{-1}.pc^2$) &    ($M_\odot$)    & ($M_\odot.yr^{-1}$)  &      (Gyr)      \\ \hline
    0          & ($0'',0''$)     &      1.83       &   $9.7\times 10^5$   & $4.5\times 10^6$  & $9.5\times 10^{-4}$  &       4.74      \\
    1          & ($0'',22''$)    &     $<0.19$     &  $<1.0\times 10^5$   & $<4.6\times 10^5$ & $6.1\times 10^{-4}$  &     $<0.75$     \\
    2          & ($-22'',22''$)  &     $<0.19$     &  $<1.0\times 10^5$   & $<4.6\times 10^5$ & $3.1\times 10^{-4}$  &     $<1.48$     \\
    3          & ($-22'',0''$)   &      0.69       &   $3.7\times 10^5$   & $1.7\times 10^6$  & $1.6\times 10^{-3}$  &       1.06      \\
    4          & ($-22'',-22''$) &      1.37       &   $7.3\times 10^5$   & $3.3\times 10^6$  & $1.2\times 10^{-3}$  &       2.75      \\
    5          & ($0'',-22''$)   &      2.40       &   $1.3\times 10^6$   & $5.8\times 10^6$  & $2.1\times 10^{-4}$  &       27.6      \\
    6          & ($22'',-22''$)  &      1.60       &   $8.5\times 10^5$   & $3.9\times 10^6$  & $4.5\times 10^{-4}$  &       8.67      \\
    7          & ($22'',0''$)    &      0.80       &   $4.2\times 10^5$   & $1.9\times 10^6$  & $3.2\times 10^{-4}$  &       5.94      \\
    8          & ($22'',22''$)   &     $<0.19$     &  $<1.0\times 10^5$   & $<4.6\times 10^5$ & $2.1\times 10^{-4}$  &     $<2.19$     \\
    9          & ($0'',44''$)    &     $<0.19$     &  $<1.0\times 10^5$   & $<4.6\times 10^5$ & $1.9\times 10^{-4}$  &     $<2.42$     \\
    10         & ($-44'',44''$)  &     $<0.19$     &  $<1.0\times 10^5$   & $<4.6\times 10^5$ & $<1.9\times 10^{-4}$ &        -        \\
    11         & ($-44'',0''$)   &      0.80       &   $4.2\times 10^5$   & $1.9\times 10^6$  & $8.6\times 10^{-6}$  &       220       \\
    12         & ($-44'',-44''$) &     $<0.19$     &  $<1.0\times 10^5$   & $<4.6\times 10^5$ & $7.0\times 10^{-4}$  &     $<0.66$     \\
    13         & ($0'',-44''$)   &      2.06       &   $1.1\times 10^6$   & $5.0\times 10^6$  & $2.1\times 10^{-4}$  &       23.8      \\
    14         & ($44'',-44''$)  &     $<0.19$     &  $<1.0\times 10^5$   & $<4.6\times 10^5$ & $1.2\times 10^{-4}$  &     $<3.83$     \\
    15         & ($44'',0''$)    &     $<0.19$     &  $<1.0\times 10^5$   & $<4.6\times 10^5$ & $1.1\times 10^{-4}$  &     $<4.18$     \\
    16         & ($44'',44''$)   &     $<0.19$     &  $<1.0\times 10^5$   & $<4.6\times 10^5$ & $4.0\times 10^{-4}$  &     $<1.15$     \\ \hline
  \end{tabular}
  \caption{\label{table:CO_S2} CO luminosities and molecular masses in the regions observed in shell S2 by \cite{Charmandaris_2000} that show evidence of star formation, either through TIR or FUV emission. $M_{H_2}$ is the molecular mass derived from the CO emission. The SFR and $t_{dep}^{mol}$ were estimated in regions of $44''$, corresponding to the beam of SEST for CO(1-0). The offset ($0'',0''$) is centred on $\alpha=13^h 24^m 35^s.4$, $\delta=-$43:08:34.9. As a result of the cirrus contamination, the SFR and $t_{dep}^{mol}$ values are subject to high uncertainties.}
\end{table*}

\begin{table*}[h]
  \centering
  \begin{tabular}{lcccc|cccc}
    \hline \hline
                    & \multicolumn{4}{c}{Shell S1 (north)} & \multicolumn{4}{c}{Shell S2 (south)} \\ \cmidrule(r){2-5}\cmidrule(r){6-9}
    Position        &     $M_{H_2}$     &    $M_{dust}$    & Tdust & $H_2$-to-dust &     $M_{H_2}$     &    $M_{dust}$    & Tdust & $H_2$-to-dust \\
                    &    ($M_\odot$)    &    ($M_\odot$)   &  (K)  &     ratio     &    ($M_\odot$)    &    ($M_\odot$)   &  (K)  &     ratio     \\ \hline
    0               & $2.8\times 10^6$  & $7.2\times 10^3$ & 23.12 &  $\sim 390$   & $4.5\times 10^6$  & $5.1\times 10^3$ & 30.00 &  $\sim 880$   \\
    1               & $1.3\times 10^6$  & $3.0\times 10^3$ & 28.16 &  $\sim 430$   & $<4.6\times 10^5$ & $3.3\times 10^3$ & 30.00 &    $<140$     \\
    2               & $1.3\times 10^6$  & $2.3\times 10^3$ & 30.00 &  $\sim 565$   & $<4.6\times 10^5$ & $1.7\times 10^3$ & 30.00 &    $<270$     \\
    3               & $1.1\times 10^6$  & $2.8\times 10^3$ & 30.00 &  $\sim 390$   & $1.7\times 10^6$  & $1.2\times 10^4$ & 28.26 &  $\sim 140$   \\
    4               & $1.3\times 10^6$  & $2.4\times 10^3$ & 30.00 &  $\sim 540$   & $3.3\times 10^6$  & $6.6\times 10^3$ & 30.00 &  $\sim 500$   \\
    5               & $1.5\times 10^6$  & $7.2\times 10^3$ & 23.21 &  $\sim 210$   & $5.8\times 10^6$  & $6.2\times 10^3$ & 21.95 &  $\sim 935$   \\
    6               & $2.1\times 10^6$  & $1.1\times 10^4$ & 18.01 &  $\sim 190$   & $3.9\times 10^6$  & $1.6\times 10^4$ & 21.22 &  $\sim 245$   \\
    7               & $<4.6\times 10^5$ & $1.9\times 10^4$ & 13.56 &     $<25$     & $1.9\times 10^6$  & $1.2\times 10^4$ & 21.24 &  $\sim 160$   \\
    8               & $2.5\times 10^5$  & $7.5\times 10^3$ & 15.39 &  $\sim 35$    & $<4.6\times 10^5$ & $7.0\times 10^3$ & 21.46 &     $<65$     \\
    9               & $4.4\times 10^5$  & $1.3\times 10^3$ & 30.00 &  $\sim 340$   & $<4.6\times 10^5$ & $1.0\times 10^3$ & 30.00 &    $<460$     \\
    10              & $<4.6\times 10^5$ &        -         &   -   &       -       & $<4.6\times 10^5$ &        -         &   -   &       -       \\
    11              & $<4.6\times 10^5$ & $2.3\times 10^3$ & 23.96 &    $<200$     & $1.9\times 10^6$  & $2.7\times 10^3$ & 14.33 &  $\sim 700$   \\
    12              & $<4.6\times 10^5$ & $1.6\times 10^3$ & 30.00 &    $<290$     & $<4.6\times 10^5$ & $1.6\times 10^4$ & 22.97 &     $<30$     \\
    13              & $1.5\times 10^6$  & $7.0\times 10^3$ & 20.04 &  $\sim 215$   & $5.0\times 10^6$  & $3.1\times 10^3$ & 24.96 &  $\sim 1615$  \\
    14              & $<4.6\times 10^5$ & $5.5\times 10^3$ & 20.45 &     $<85$     & $<4.6\times 10^5$ & $1.0\times 10^4$ & 17.98 &     $<45$     \\
    15              & $<4.6\times 10^5$ &        -         &   -   &       -       & $<4.6\times 10^5$ & $1.1\times 10^4$ & 17.66 &     $<40$     \\
    16              & $<4.6\times 10^5$ &        -         &   -   &       -       & $<4.6\times 10^5$ & $5.8\times 10^3$ & 25.13 &     $<80$     \\ \hline
  \end{tabular}
  \caption{\label{table:gas-dust} Molecular gas and dust masses in the regions observed in shell S1 (\emph{left} and shell S2 (\emph{right}). We also indicate the dust temperature fitted by the modified black-body. When possible, a molecular gas-to-dust ratio is calculated. The dust mass is very sensitive to IR background subtraction uncertainties.}
\end{table*}

\begin{acknowledgements}
      We thank the anonymous referee for his/her useful remarks. We also thank Raffaella Morganti for providing the ATCA data of the radio continuum, and Katherine Alatalo and Sperello di Serego Alighieri for constructive comments on the article. \\

   This paper makes use of the following ALMA data: ADS/JAO.ALMA$\#$2011.0.00454.S. ALMA is a partnership of ESO (representing its member states), NSF (USA) and NINS (Japan), together with NRC (Canada) and NSC and ASIAA (Taiwan) and KASI (Republic of Korea), in cooperation with the Republic of Chile. The Joint ALMA Observatory is operated by ESO, AUI/NRAO and NAOJ.\\

   The Australia Telescope Compact Array is part of the Australia Telescope National Facility, which is funded by the Commonwealth of Australia for operation as a National Facility managed by CSIRO. \\

   This research has made use of the NASA/IPAC Extragalactic Database (NED), which is operated by the Jet Propulsion Laboratory, California Institute of Technology, under contract with the National Aeronautics and Space Administration. \\

   GALEX (Galaxy Evolution Explorer) is a NASA Small Explorer, launched in 2003 April. We gratefully acknowledge NASA’s support for construction, operation, and science analysis for the GALEX mission, developed in cooperation with the Centre National d’Etudes Spatiales of France and the Korean Ministry of Science and Technology. \\

   Herschel is an ESA space observatory with science instruments provided by European-led Principal Investigator consortia and with important participation from NASA. \\

   F.C. acknowledges the European Research Council for the Advanced Grant Program Number 267399-Momentum.
\end{acknowledgements}

\bibliography{Biblio}
\bibliographystyle{aa}

\end{document}